\documentclass[
twocolumn,
superscriptaddress,
amsmath,amssymb,
]{revtex4-2}
\usepackage{graphicx}
\usepackage{hyperref}
\usepackage{dcolumn}
\usepackage{bm}
\usepackage{epstopdf}
\usepackage{romannum}
\usepackage{csquotes}
\usepackage[dvipsnames]{xcolor}
\usepackage{hyperref}
\usepackage{float}

\begin{document}
	
\preprint{}
	
\title{Increase of critical current density in FeSe superconductor by strain effect}
	
\author{Han Luo}
\affiliation{School of Physics, Southeast University, Nanjing 211189, China}
\author{Xinyue Wang}
\affiliation{School of Physics, Southeast University, Nanjing 211189, China}
\author{Xin Zhou}
\affiliation{School of Physics, Southeast University, Nanjing 211189, China}
\author{Longfei Sun}
\affiliation{School of Physics, Southeast University, Nanjing 211189, China}
\author{Mengqin Liu}
\affiliation{School of Physics, Southeast University, Nanjing 211189, China}
\author{Ran Guo}
\affiliation{School of Physics, Southeast University, Nanjing 211189, China}
\author{Sheng Li}
\affiliation{School of Physics, Southeast University, Nanjing 211189, China}
\author{Yue Sun}
\email{Corresponding author: sunyue@seu.edu.cn}
\affiliation{School of Physics, Southeast University, Nanjing 211189, China}
\author{and Zhixiang Shi}
\email{Corresponding author: zxshi@seu.edu.cn}
\affiliation{School of Physics, Southeast University, Nanjing 211189, China}

\begin{abstract}
		
\textbf{}
Conventional $J_c$-enhancement methods like doping and irradiation often introduce extrinsic elements or defects, altering intrinsic properties. Here, we report a significant $J_c$ enhancement in FeSe single crystals through compressive strain applied using a glass-fiber-reinforced plastic substrate with anisotropic thermal contraction during cooling. Under zero field at 2 K, $J_{\text{c}}$ increases by a factor of $\sim$4 from $\sim 2.3 \times 10^{4}$ to $\sim 8.7 \times 10^{4}$ A cm$^{-2}$; at 5 T, it achieves an order-of-magnitude enhancement, rising from $\sim 1.0 \times 10^{3}$ to $\sim 1.0 \times 10^{4}$ A cm$^{-2}$. Analysis based on the Dew-Hughes model of the \(f_p(h)\) relationship shows that strain strengthens vortex pinning, and shifts the pinning mechanism from point-like pinning to combined point and surface pinnings. This work offers an effective method to enhance FeSe's current-carrying limitation, deepens understanding of iron-based superconductors' pinning mechanisms, and highlights strain engineering's potential for optimizing superconducting performance.
\end{abstract}
	
	
\maketitle

\section{introduction}

\textbf{}
The critical current density \(J_c\), which represents the upper limit of a superconductor's zero-resistance current-carrying capacity, directly determines the performance boundaries in applications such as high-field magnets and power transmission \cite{li2019giant,blatter1994vortices,hosono2018recent}. A high \(J_c\) enables superconducting materials to transport larger currents, making it essential for superconducting power transmission systems, superconducting magnets, and superconducting electronic devices \cite{larbalestier2001high,larbalestier2014isotropic}. However, the widespread application of superconductors is critically hampered by limitations in simultaneously achieving high \(J_c\) and other essential properties in practical materials.  While copper oxide superconductors currently hold the record for the highest \(J_c\) values, their strong anisotropy, a small critical grain boundary angle \(\theta_b\), and high manufacturing costs pose significant challenges for industrial applications \cite{li2019giant,hosono2018recent}. In contrast, iron-based superconductors, combining a high upper critical field with low anisotropy \cite{hosono2018recent,hunte2008two,moll2010high,katase2011advantageous,li2011films,si2013high,haindl2014thin,zhang2014realization,hanisch2019fe}, are promising for high-field magnets and loss-free power transmission.

 FeSe, the structurally simplest parent of the iron-based family \cite{sun2017effects,sun2015critical}, possesses a layered lattice that exhibits a superconducting transition at 9\,K under ambient pressure without chemical doping \cite{mcqueen2009tetragonal}. Its easy cleavability and low toxicity render FeSe an ideal platform for investigations of vortex dynamics \cite{terashima2024transport}, where understanding and controlling flux pinning is paramount to achieving high \(J_c\). Nevertheless, the intrinsic critical current density of FeSe single crystals remains only $\sim$\,$2.3 \times 10^{4}$\,A\,cm$^{-2}$ at 2\,K under self-field \cite{meng2022significant,sun2017effects,sun2015critical,sun2016effect,sun2015enhancement,sun2016effect}, far below the practical threshold and thus constituting a key limiting factor for applications. Previous attempts to enhance \(J_c\) via chemical intercalation, aliovalent doping, or ion irradiation have indeed achieved substantial improvements. For example, sulfur doping introduces point-pinning center, and elevates \(J_c\) from \(\sim 4 \times 10^4\)\,A cm\(^{-2}\) to \(\sim 6 \times 10^4\)\,A cm\(^{-2}\) at 4\,K \cite{sun2016effect}. Similarly, \( H^+ \) irradiation introduces extra point pinning centers in FeSe single crystals, increasing the \(J_c\) from \(\sim 3 \times 10^4\) to \(\sim 8 \times 10^4\)\,A cm\(^{-2}\) under zero field, which is an enhancement of over two times \cite{sun2015enhancement}. Uranium irradiation generates \(c\)-axis-aligned columnar defects, raising \(J_c\) at 2\,K from \(4 \times 10^4\)\,A cm\(^{-2}\) to \(2 \times 10^5\)\,A cm\(^{-2}\) and shifting the dominant vortex pinning mechanism from \(\delta T_c\) pinning to \(\delta l\) pinning \cite{sun2017effects}. Critically, a common drawback of these approaches is the inevitable introduction of extrinsic elements or radiation-induced defects, which usually suppresses the value of \(T_c\) and alters the stoichiometry and electronic structure. Consequently, achieving substantial \(J_c\) enhancement in FeSe via physical methods under ambient pressure remains a fundamental challenge.

Here, we report the strain-engineered enhancement of \(J_c\) in FeSe. By exploiting the anisotropic thermal contraction of a glass-fiber-reinforced plastic substrate (GFRP) during cooling, compressive strain is applied to FeSe single crystals. Measurements reveal that applied strain induces a pronounced enhancement in the \(J_c\), yielding an approximately fourfold increase at zero field, and an order-of-magnitude enhancement under a 5\,T magnetic field. The pinning mechanism transforms from normal point pinning to a coexistence of normal point and surface pinning following strain application. This substrate-mediated strain-engineering strategy not only offers an effective route to surpass the current-carrying limiting factor of FeSe, but also highlights the universal potential of strain control for optimizing superconducting performance.
	
\begin{figure*}[htbp]
	\centering
	\includegraphics[width=1\linewidth]{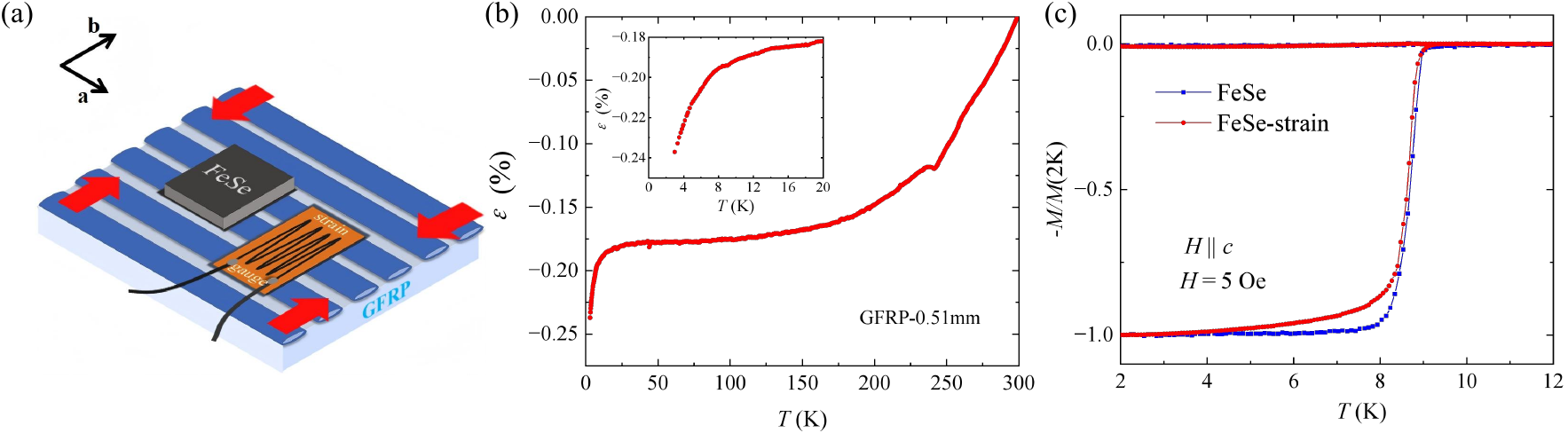}
	\caption{(a) Schematic illustration of the contraction of the GFRP substrate and the strain measurement setup. The FeSe sample and the strain gauge were bonded using epoxy resin, with the tetragonal [100] direction of the FeSe crystal oriented parallel to the fibers of the GFRP substrate, and the sensitive grid of the strain gauge oriented perpendicular to the fibers. During the cooling process, the GFRP substrate contracts in the direction indicated by the red arrow. (b) Strain values of the 0.51 mm thick GFRP substrate as a function of temperature. The inset shows an enlarged view of the strain from 0 to 20 K. (c) Temperature dependence of the ZFC and FC magnetization of FeSe and FeSe-strain single crystals.}
	\label{Fig1}
\end{figure*}

\section{experimental details}

Pristine FeSe single crystals were obtained by the chemical vapor transport growth \cite{sun2015critical,sun2016electron,hou2024bulk}. To apply strain, a FeSe single crystal (1087\,×\,1048\,×\,75\,\textmu m\textsuperscript{3}) was cleaved and then glued onto a glass-fiber-reinforced plastic (GFRP) substrate (3730\,×\,1826\,×\,466\,\textmu m\textsuperscript{3}) using epoxy (STYCAST 2850\,FT BK 0.5\,K, with Catalyst 24\,LV), ensuring that the tetragonal [100] direction of the crystal was made perpendicular to the fiber direction. GFRP consists of two main components: glass fibers, which provide strength and support; synthetic resin, which acts as a binder to hold the fibers together. During the cooling process, the GFRP substrate contracts more in the direction perpendicular to the fibers than in the parallel direction \cite{he2017dichotomy}. To effectively transfer the compressive strain to FeSe, the epoxy resin was thinly and evenly applied to the FeSe. To accurately determine the magnitude of the compressive strain applied to the FeSe crystal along the tetragonal [010] direction, an in-situ calibration was conducted using a standard resistive strain gauge (CFLA-1-350-11-6LTA-F). The strain gauge was firmly bonded to the same GFRP substrate as the FeSe crystal, following the identical epoxy bonding procedure. This combined substrate--strain-gauge assembly was installed in a Physical Property Measurement System (PPMS-9\,T) and underwent the same temperature-cycling protocol as that used in the superconducting measurements. The non-magnetic property of GFRP makes it suitable for magnetic measurements \cite{SupplMaterial1}. Magnetic measurements were performed using the Magnetic Property Measurement System (MPMS‐XL5, Quantum Design). The magnetic field sweep rate was set to 120\,Oe\,s$^{-1}$ for magnetic hysteresis loops (MHLs).

	\begin{figure*}[htbp]
	\centering
	\includegraphics[width=15cm]{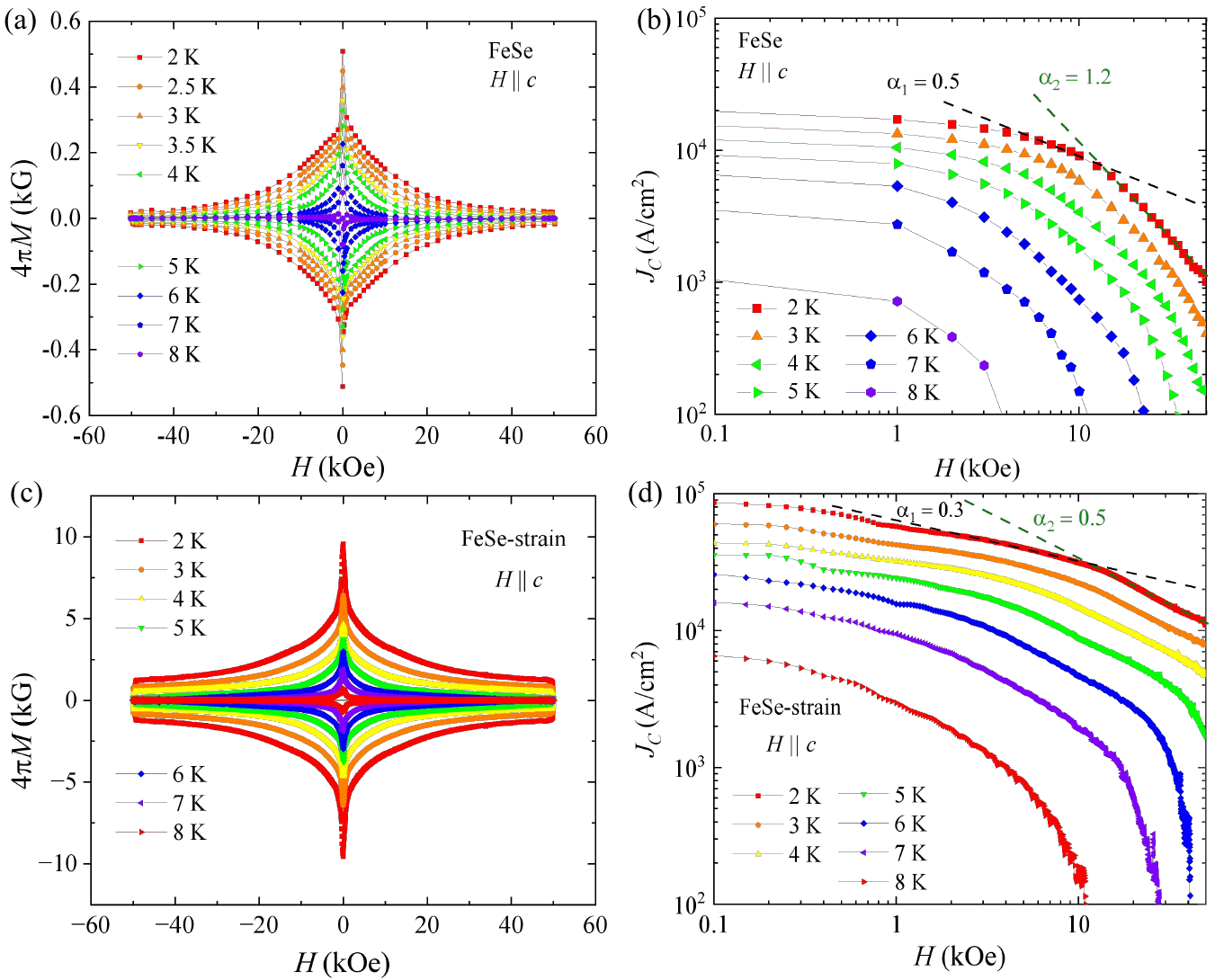}
	\caption{The temperature dependence of MHLs for \(\textit{H}\,||\,\textit{c}\) and the magnetic field dependence of the \(J_{\text{c}}\) derived from the Bean model. (\,a\,) and (\,b\,) correspond to FeSe, while\,(\,c\,) and\,(\,d\,) correspond to FeSe-strain. In panels (b) and (d), the dashed lines represent the exponent \(\alpha\) obtained from fitting the critical current density to the power-law relation \(J_{\text{c}} \propto H^{-\alpha}\). The black dashed line indicates the exponent in the range of 5\,--\,10\,kOe, while the green dashed line corresponds to the exponent above 10\,kOe.}
	\label{Fig2}
\end{figure*}

\section{results and discussion}

Fig.\,1(a) illustrates the experimental setup designed to transmit compressive strain to the FeSe crystal. The crystal and a resistive strain gauge were bonded to the GFRP substrate using epoxy resin. The tetragonal [100] direction of the FeSe crystal was aligned parallel to the substrate's fibers, while the sensitive grid of the strain gauge was oriented perpendicular to them. During cooling, the GFRP substrate contracts predominantly along the direction indicated by the red arrow. In this configuration, a strain gauge is a sensor that converts the deformation of a material into changes in resistance \cite{mo2024cryogenic,chu2012divergent,hosoi2016nematic}. In the context of iron-based superconductors, strain gauges have been widely employed to characterize anisotropic strain, particularly in studies of electronic nematicity and structural transitions \cite{kuo2013measurement,kuo2016ubiquitous,palmstrom2017critical}. For instance, in FeSe and related compounds, strain gauges have been used to measure the transmission efficiency of strain from substrates to crystals \cite{tanatar2016origin}. To measure the strain experienced by FeSe due to the contraction of 0.51\,mm-thick GFRP at low temperatures, a resistive strain gauge was employed. As the temperature decreases, the GFRP contracts in the direction perpendicular to the fibers, causing the sensitive grid of the strain gauge attached to it to deform. This deformation leads to a change in the resistance of the strain gauge. The change in resistance is proportional to the strain  \(\epsilon\) via \(\Delta R/R = K\epsilon\), where \(K\) is the gauge factor of the strain gauge. This relationship allows for the determination of the strain value. Fig.\,1(b) shows the strain of the 0.51\,mm-thick GFRP as a function of temperature. As the temperature decreases, the contraction of the GFRP increases. It contracts rapidly during the cooling process down to 250\,K. Below 250\,K, the contraction rate becomes relatively slower. The fastest contraction occurs when the temperature drops from 10\,K to 3\,K. At 3\,K, the strain in the GFRP is approximately \(-0.24\%\).

Fig.\,1(c) illustrates the temperature dependence of field cooling (FC) and zero FC (ZFC) magnetization at 5 Oe for FeSe and FeSe-strain single crystals. The superconducting transition temperature \(T_{\text{c}}\) of FeSe-strain remains essentially unchanged at {9}\,K, comparable to that of pristine FeSe. This result demonstrates that the strain effect has no influence on the \(T_c\) of FeSe. Hydrostatic pressure has a significant effect on the \(T_c\) of FeSe \cite{sun2016dome}. This is because under hydrostatic pressure, the FeSe lattice is uniformly compressed, which alters the band structure and enhances spin fluctuations, thus significantly increasing \(T_c\) \cite{matsuura2017maximizing}. In contrast, the compressive strain is applied in the \(ab\) plane and mainly tunes the nematicity and the associated orbital degrees of freedom. Its effect on the overall Fermi-surface volume and the enhancement of spin fluctuations is limited. For instance, as demonstrated in a recent study on FeSe under in-plane uniaxial strain along the [110] direction, a compressive strain of about \(0.2\%\) induces only a minimal shift in \(T_c\) of approximately 0.2\,K \cite{liu2025evolution}.

In Fig.\,2(a), the MHLs of FeSe and FeSe-strain at different temperatures are measured and compared. The MHLs for both crystals are nearly symmetric, reflecting the dominance of bulk pinning. Unlike many other iron-based superconductors \cite{prozorov2008vortex,haberkorn2011strong,salem2010flux,xing2020vortex,xing2019reemergence}, these loops show no evidence of a second magnetization peak, and \(M\) displays a monotonic decrease with increasing \(H\). By selecting appropriate normalization parameters \(M\) and \(H\), MHLs measured at different temperatures can be normalized onto a single curve if one pinning mechanism is dominant \cite{perkins1995implications,oussena1994magnetization,dewhurst1996determination,dewhurst2000separation}. When multiple pinning mechanisms coexist, the normalized curves will not overlap, thereby intuitively reflecting the variation in pinning mechanisms.  The maximum magnetization value\,(\textit{M}$^*$)\,are selected and the irreversibility field\,(\textit{H}$^*$)\,is determined by extrapolating \(J_{\text{c}}\) to zero in the \(J_{\text{c}}^{1/2}\)-\(H\) curve \cite{sun2015critical,sun2017effects,SupplMaterial2}. As shown in Figs.\,3\,(a) and 3\,(b), the normalized MHLs for FeSe measured at different temperatures scaled well, indicating that the FeSe crystal is dominated by a single pinning mechanism. In contrast, the normalized MHLs for FeSe-strain are altered and do not fully overlap across temperatures. Fig.\,3(c) compares the scaling MHLs of FeSe and FeSe-strain at\,5\,K. The difference is due to the variation in pinning mechanisms caused by the application of compressive strain, which will be discussed in detail later.

 \begin{figure*}[htbp]
 	\centering
 	
 	\includegraphics[width=1\linewidth]{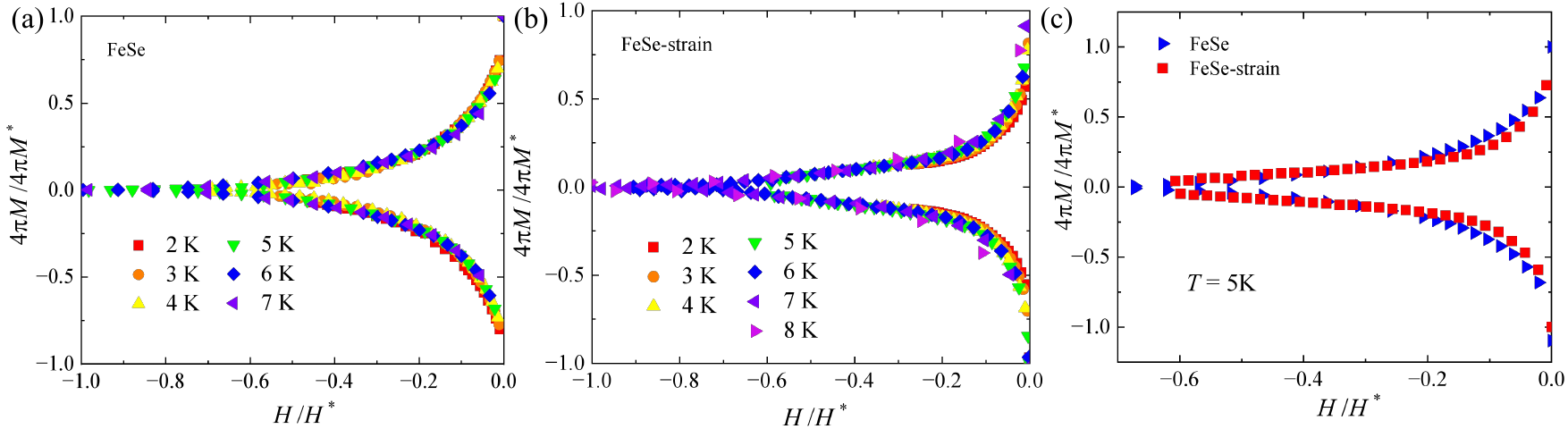}
 	\caption{Scaled MHLs for (a)\,FeSe and (b)\,FeSe-strain, respectively, at different temperatures. (c)\,Scaled MHLs at 5\,K for the pristine and strained crystals.}
 	\label{Fig3}
 \end{figure*}
 
 \begin{figure}[htbp]
 	\centering
 	\includegraphics[width=1\linewidth]{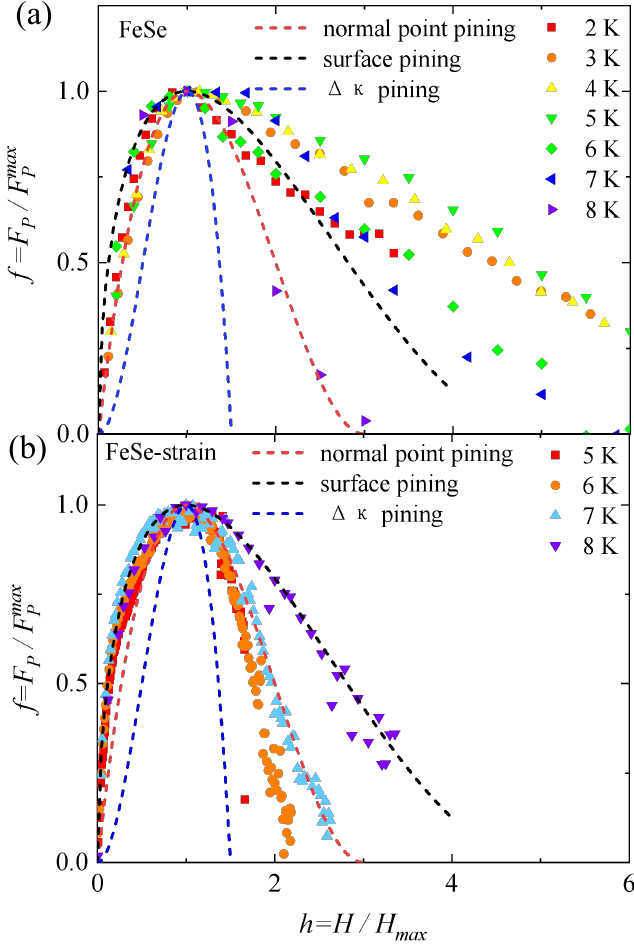}
 	\caption{The temperature variation of the normalized vortex pinning force (\(f = F_{\text{p}} / F_{\text{pmax}}\)) and reduced field (\(h = H/H_{\text{max}}\)) for FeSe and FeSe-strain.}
 	\label{fig:temp_variation}
 \end{figure}
 
 The critical current density \(J_{\text{c}}\) can be extracted from the MHLs using the extended Bean model \cite{bean1964tviagnetization}, which is given by:
\begin{equation*}
	J_{\text{c}} = \frac{20 \Delta M}{a(1 - a/3b)}.
\end{equation*} Here, \(\Delta M\) (\(\text{emu cm}^{-3}\)) denotes the difference between the magnetization during the upward and downward field sweeps, while \(a\) and \(b\) dcorrespond to the width and length of the crystals, respectively. It should be noted that the pre-factor 20 in this expression is dimensionless when consistent magnetic units are employed, as emphasized in recent work on unit conversion in Bean's model \cite{polichetti2025equation}. Figs.\,2(b) and\,2(d) present the \(J_{\text{c}}\) of FeSe and FeSe-strain at various temperatures. At zero field and 2\, K, the \(J_c\) of pristine FeSe is $\sim$\,\(2.3 \times 10^4\) A cm\textsuperscript{$-2$}. With increasing the magnetic field to 5\, T, \(J_c\) decreases to \(1.0 \times 10^3\) A cm\textsuperscript{$-2$}. At zero field, the \(J_c\) of FeSe-strain reaches approximately \(8.7 \times 10^4\) A cm\textsuperscript{$-2$}, representing an approximately fourfold increase compared to its unstrained state. Under a 5\, T magnetic field, the \(J_c\) of FeSe-strain is \(1.0 \times 10^4\) A cm\textsuperscript{$-2$}, an enhancement by one order of magnitude over that of unstrained FeSe single crystals. This substantial improvement in \(J_c\) at high fields originates from strain-induced strengthening of flux pinning, enabling FeSe to retain a significantly higher dissipation-free current capacity under these conditions. Collectively, these results demonstrate that the strain effect is effective in enhancing the superconducting performance of FeSe.

\(J_c\) for both crystals exhibits minimal variation below 1\, kOe. Above 5\, kOe, \(J_c\) decays according to a power law, \(H^{-\alpha}\). For FeSe, in the magnetic field range of 5\,-\,10\, kOe, the power-law exponent \(\alpha\) is approximately 0.5. This is attributed to the strong pinning by sparse nanometer-sized defects, similar to the case of YBCO films \cite{van2002strong,van2010flux}. In high magnetic fields above 10\, kOe, the decay rate of \(J_c\) significantly increases to \(\alpha \sim 1.2\). Since the pinning force \(F_p = \mu_0 H \cdot J_c\), the fact that \(F_p\) remains constant implies that \(J_c\) decreases proportionally to \(H^{-1}\) with increasing magnetic field \cite{krusin1996accommodation}. In FeSe, high magnetic fields cause the limited strong pinning centers to become fully occupied by flux lines. This saturation of the pinning force results in a rapid suppression of \(J_c\), which consequently exhibits an inverse proportionality to \(H\). In contrast, FeSe-strain shows a more gradual decay in the 5\,-\,10\,kOe magnetic field range, with a power-law exponent \(\alpha \sim 0.3\). Computational studies employing large-scale time-dependent Ginzburg-Landau theory have demonstrated that it is due to the pinning from spherical defects \cite{park2020effects,taen2015critical,taen2012enhancement, willa2017strong}. Above 10\,kOe, the power-law exponent \(\alpha\) for FeSe-strain remains at 0.5, significantly lower than the \(\alpha \sim 1.2\) observed for FeSe, indicating that strain creates a new pinning center. After applying strain, the field dependence of \(J_c\) is modified. The power-law exponent shifts from \(H^{-0.5}\) to \(H^{-0.3}\) at low magnetic fields. Similarly, the field dependence of \(J_c\) undergoes pronounced attenuation at high magnetic fields, reducing from \(H^{-1.2}\) to \(H^{-0.5}\). Strain effectively maintains the binding ability of pinning centers on vortex lines, suppressing the weakening of pinning barriers due to vortex-vortex interactions at high magnetic fields, thereby avoiding a sharp drop in \(J_c\).

 Various experimental methods have been proven to effectively enhance the \(J_c\) of FeSe. For instance, sulfur (S) doping-induced lattice distortion can reduce the size of magnetic flux bundles and increase effective pinning energy. In the FeSe\textsubscript{0.86}S\textsubscript{0.14} single crystals, the \(J_c\) at 2\, K and self-field increases to approximately \(6 \times 10^4\) A cm\textsuperscript{$-2$}, compared with the \(J_c\) of about \(4 \times 10^4\) A cm\textsuperscript{$-2$} in the original FeSe single crystals \cite{sun2016effect}. Ion irradiation is another approach that can be employed to enhance the \( J_c \) of FeSe. By appropriately selecting ion types and energies for irradiation, various defects can be introduced into superconducting materials, which in turn enables the control of pinning defects. For instance, point defects were successfully introduced into FeSe single crystals through 3\,MeV \( H^+ \) irradiation \cite{sun2015enhancement}. Consequently, the \(J_c\) increased to approximately \( 8 \times 10^4 \)\,A cm\(^{-2}\), more than doubling the original value. Furthermore, uranium irradiation can introduce columnar defects along the \textit{c} axis in FeSe single crystals \cite{sun2017effects}. As a result, the \( J_c \) at 2\,K was enhanced to \( 2 \times 10^5 \)\,A cm\(^{-2}\), which is more than five times higher than the original \( 4 \times 10^4 \)\,A cm\(^{-2}\). Compared with the situation where \(T_c\) drops dramatically from 9.2\,K to 5\,K after irradiation by uranium with \(B_{\Phi} = 16\,\text{T}\), strain tuning achieves a significant increase in \(J_c\) while maintaining \(T_c\) almost unchanged, providing a new approach for optimizing the performance of FeSe superconducting materials.
 
 To achieve a more thorough comprehension of the vortex pinning mechanism, Figs.\,4(a) and\,4(b) display the temperature dependence of the normalized vortex pinning force (\(f = F_{\text{p}} / F_{\text{pmax}}\)) and reduced field (\(h = H/H_{\text{max}}\)) for FeSe and FeSe-strain. Here, \(H_{\text{max}}\) indicates the magnetic field peak, whereas \(F_{\text{pmax}}\) represents the maximum pinning force. The \(F_{\text{p}}\) determines the motion and positioning of vortices in a magnetic field, and it was calculated using \(F_{\text{p}} = \mu_0 H \times J_{\text{c}}\). Within the Dew-Hughes model \cite{dew1974flux}, there is a specific functional relationship between \(f\) and \(h\), given by \(f_{\text{p}} \propto h_{\text{p}}(1 - h)^q\), where the parameters \(p\) and \(q\) are determined by the pinning mechanism. The main pinning mechanism includes \(\Delta \kappa\) pinning, normal point pinning, and surface pinning. Within a particular temperature range, if \(h\) meets any of the following equations, it shows that the corresponding pinning mechanism is dominant.
\begin{align}
	&f_{\text{p}} = \frac{9}{4}h(1 - h/3)^2 \quad \text{for normal point pinning}, \tag{1} \\
	&f_{\text{p}} = \frac{25}{16}h^{1/2}(1 - h/5)^2 \quad \text{for surface pinning}, \tag{2} \\
	&f_{\text{p}} = 3h^2(1 - 2h/3) \quad \text{for } \Delta \kappa \text{ pinning}. \tag{3}
\end{align}
For high-\(T_c\) superconductors, flux creep is highly pronounced, particularly under high fields, and exhibits both temperature and field dependence. Consequently, the pinning force \(F_p\) usually cannot be scaled at high fields, as has been observed in cuprates \cite{higuchi1999comparative,klein1994peak,civale1991weak} and in several iron-based superconductors (IBSs) \cite{li2014giant,fang2011doping}. Although the \(T_c\) of FeSe is relatively low, its flux-creep rate is comparable to that of other high-\(T_c\) superconductors \cite{sun2015critical,sun2013magnetic}. Owing to this failure of scaling at high fields, such data cannot be directly compared with the theoretical models \cite{huang2022magnetization,pan2021anisotropic}. Therefore, the analysis is restricted to the low-field regime (\(h<1\)), where flux creep has minimal influence. In this regime, the scaled data for pristine FeSe can be reasonably fitted by the normal point-pinning model, as shown in Fig.\,4\,(a). In contrast, the strained FeSe data lie between the theoretical curves for normal point pinning and surface pinning. These results are further supported by the slope change observed in the field dependence of \(J_c\), as shown in Fig.\,2.

In pristine FeSe, \(f_{\mathrm{P}}\) between 2\,-\,6\,K can be well fitted with equation\,(1), indicating that normal point pinning is the dominant pinning mechanism. The pinning centers for normal point pinning are located within the superconductor, such as at impurities or defects, around which flux lines distribute, and the \(J_c\) of the superconductor is mainly influenced by these internal pinning centers. For FeSe-strain, between 2\,-\,4\,K, the \(F_{\text{P}}\)\,-\,\(H\) curves diverge, making it hard to determine \(F_{\text{pmax}}\) \cite{SupplMaterial3}. Hence, Fig.\,3(b) presents the \(f\)-\(h\) results at 5\,-\,8\,K. Within 5\,-\,8\,K, the \(f_{\mathrm{P}}\) pinning mechanism of FeSe\,-\,strain exhibits intermediate characteristics between normal point pinning and surface pinning, suggesting the coexistence of both pinning centers. Normal point pinning refers to the internal pinning centers’ dominant role in flux line pinning, while the surface pinning implies the surface or interface-related pinning is dominant. This transition in pinning mechanism reveals the physical origin underlying the previously observed reduction in the power-law exponent \(\alpha\) of \(J_{\text{c}} \propto H^{-\alpha}\). Specifically, the enhanced surface pinning induces a significant decrease in \(\alpha\) from \(\sim 1.2\) to \(\sim 0.5\).
Future studies could employ high-resolution transmission electron microscopy (HRTEM) and scanning tunneling microscopy (STM) to directly investigate the types and spatial distribution of defects introduced in strained FeSe crystals. This approach would help clarify the intrinsic nature of strain-induced pinning centers and elucidate the mechanisms underlying their enhanced vortex-pinning capability. 
	
\section{conclusion}
In summary, compressive strain was applied to FeSe single crystals using an anisotropically contracting GFRP substrate. While the $T_c$ remains unchanged, the $J_c$ is significantly enhanced. At 2\,K and zero field, $J_c$ increases approximately fourfold\,, and achieves an order-of-magnitude enhancement under a 5\,T magnetic field. Analysis shows that the strain reduces the decay rate of $J_c$, confirming the enhanced flux pinning strength. This mitigates the weakening of the pinning potential barrier by vortex interactions at high fields, preventing the sharp $J_c$ decline. Dew-Hughes model analysis further reveals a coexistence of normal point pinning and surface pinning in FeSe-strain. These results demonstrate that strain effectively enhances flux pinning in FeSe by modifying its structural and defect characteristics, offering a promising pathway for optimizing FeSe and related iron-based superconductors.
\acknowledgements	
This work was partly supported by the National Key R\&D Program of China (Grants No.\, 2024YFA1408400) and the National Natural Science Foundation of China (Grants No.\, 12374136, No.\, 12374135, No.\, U24A2068, No.\, 12204487).                       

H.L. and X.W. contributed equally to this work.

\bibliographystyle{unsrt}
\bibliography{ref}
	
\end{document}